\documentstyle[aps]{revtex}
\input{epsf.sty}\input{psfig.sty}
\begin{document}
\title{Direct Observation of Tunnelling in KDP Using Neutron Compton Scattering}
\author{G.F.Reiter}
\address{Physics Department and Texas
Center for Superconductivity\\
University of Houston, Houston,  Texas, USA}
\author{J.  Mayers}
 \address{ Rutherford
-Appleton Laboratory, Didcot, England}
\author{P.  Platzman}\address{Bell Labs, Murray Hill, New Jersey, USA  }
 \maketitle
 \begin{abstract}
  Neutron Compton Scattering
measurements presented here of the momentum distribution of
hydrogen in $KH_2PO_4$ (KDP) just above and well below the
ferroelectric transition temperature show clearly that the proton
is coherent over both sites in the in the high temperature phase,
a result that invalidates the commonly accepted order-disorder
picture of the transition.
 The Born-Oppenheimer potential for the hydrogen, extracted directly from
 data for the first time,
is consistent with neutron diffraction data,  and the vibrational
spectrum is in substantial agreement with infrared absorption
measurements. The measurements are sensitive enough to detect the
effect of surrounding ligands on the hydrogen bond, and can be
used to study the systematic effect of the variation of these
ligands.
\end{abstract}
\vskip 12pts PACS numbers, 61.12-q, 64.60-i,77.80-c \vskip .5in
 It is known that in KDP\cite{rn}

 each proton is equally likely to be in either of two
positions in the hydrogen bond above the ferroelectric phase
transition at Tc=124K, where the structure is tetragonal, and is
nearly entirely in one site below Tc, where the structure is
orthorhombic. What has not been known, and which has been a
subject of controversy since Blinc\cite{rb}  proposed the idea in
1973, is whether the protons are tunnelling between the two
equivalent sites or are localized in one or the other of the two
sites and simply disordered above the transition. While the
tunnelling model was used to describe the data for some time after
it was proposed, Raman scattering data\cite{tut} that seemed to
show that the
 symmetry of the surroundings of the proton did not change above and below the transition, together
 with subsequent neutron scattering results\cite{iy}, led to the consensus that the proton was self trapped
 in one or the other of its equivalent positions, and jumped from position to position
through phonon assisted tunnelling. If such a picture were
correct, the proton would be in one site
 or the other in the high temperature phase for times much longer than the time scale of our
  experiment($10^{-15}$ sec), not simultaneously in both.
The transition would
 be an order-disorder type in which the populations of the two sites became unequal below the ordering
 temperature. As pointed out by Reiter and Silver\cite{rs}, neutron Compton scattering provides a means to
 distinguish between the tunnelling and the order disorder model. If the latter were correct, NCS
 should see only a small change in the
momentum distribution. This would be  due to whatever changes
occurred in going through the transition to the potential of
individual wells in which the proton was trapped. By contrast, if
the particle were tunnelling and then became trapped, we should
see a narrowing of the momentum distribution in the tunnelling
phase, together with an oscillation due to the coherent
interference of the proton in the two sites. We show here that the
distributions at temperatures above and below the ferroelectric
transition are, in fact,  dramatically different, and consistent
with the proton coherently occupying the two sites. Since the bond
is symmetric above the transition, one can construct the effective
Born-Oppenheimer potential directly (without any model) from the
measurement of the momentum distribution\cite{rs}. This is as far
as we know, the first such measurement of a Born-Oppenheimer
potential in any system. It
 yields a double well potential with parameters consistent with neutron crystallography and incoherent
  neutron scattering at lower scattering energies, and yields vibration frequencies consistent with
  infrared measurements.
  The experiments are done on the electron volt
spectrometer, EVS, at ISIS, the pulsed neutron source at the
Rutherford Laboratory. This sort of source is needed to provide
high energy neutrons(5-10 ev) for which the energy transfer is
sufficiently large compared to the characteristic energies of the
system that the scattering is given accurately by the
impulse\cite{pmp} approximation limit.  The scattering at these
energies is entirely incoherent, each particle scattering
independently. $S_M(\vec{q},\omega$), the scattering function for
a particle of mass M,  is related to the momentum distribution of
the particle n($\vec{p}$) in this limit by the relation

\begin{equation}
S_M(\vec{q},\omega) = \int n(\vec{p})\delta(\omega-{\hbar
q^2\over2M}-{\vec{p}.\vec q\over M}) d\vec{p} \label{sqw}
\end{equation}
where $\hbar\omega$ is the energy transfer, M is the mass of the
proton, and q=$|\vec{q}|$ is the magnitude of the  wave-vector
transfer.

The small mass of the proton leads to a broad distribution in
energy of the scattered neutrons, centered at  ${\hbar^2q^2\over
2M}$, that is well separated from the scattering from  the heavier
ions such as oxygen, which appear as nearly elastic contributions.
This, together with its large incoherent cross-section, make it an
ideal candidate for these measurements.
 $n(\vec{p})$, the probability of observing the proton with momentum
$\vec{p}$, for simple one particle systems  in their ground state,
is the square of the absolute value of the fourier transform of
the spatial wave-function. The experimental data is fit in a model
free way using a series expansion that allows one to reconstruct
n(p)directly from the fitted coefficients.  We represent
$S_M(\vec{q},\omega)$ as ${M\over q} J(\hat {q}, y)$  where
y=${M\over q}(\omega-{q^2\over 2M})$, and expand $J(\hat {q}, y)$
as
\begin{equation}
J(\hat q, y) = {e^{-y^2}\over \pi^{1\over 2}} \sum\limits_{n,l,m}
a_{n,l,m} H_{2n+l}(y)Y_{lm}(\hat q) \label{exp}
 \end{equation}
 where the $H_n(y)$ are Hermite polynomials and the $Y_{lm}$ are
 spherical harmonics. This series is truncated at some order(2n+l=10) in this case)
 , convolved with the
 instrumental resolution function and then least squares fit
 to the data. The coefficients $a_{n,l,m}$  then  determine the
 measured $n(\vec p)$ directly as a series in Laguerre polynomials and spherical
 harmonics\cite{rs,rmn} The procedure is a smoothing operation, which works with noisy data,
   and which also allows for the inclusion of small corrections to the impulse
 approximation\cite{rmn,vs}. The errors in the measured $n(\vec p)$ are
 determined by the uncertainty in the the measured
  coefficients,through their correlation matrix, which is
 calculated by the fitting program. For an expanded discussion of the
 procedure and errors see Ref 7.
  In fitting the data, we take the z axis to be along the bond axis,
the x axis to be the c axis of the crystal, perpendicular to both
bonds, and fit the data with the sum of two terms, one rotated by
90 degrees about the x axis  from the other. The measurement
provides the complete 3-D momentum distribution. We show in Fig. 1
a cut along the $p_x-p_z$ plane for the momentum distribution of a
single bond at two temperatures, one far below and one just above
the transition. The measurements were taken with exactly the same
experimental configuration, and analyzed with exactly the same
terms kept in the series expansion. If the order-disorder picture
of the transition had been correct, we would expect to see small
changes in the momentum distribution on going through the
transition, as the proton would remain localized in one or the
other equivalent sites, and the confining potential at that site
would be only slightly modified. In fact we see dramatic changes
in the width and the shape of the momentum distribution. While
there is little change in the direction normal to the bond, the
distribution along the bond changes qualitatively.

\begin{figure}[h]
\hspace{.25in} \psfig{figure=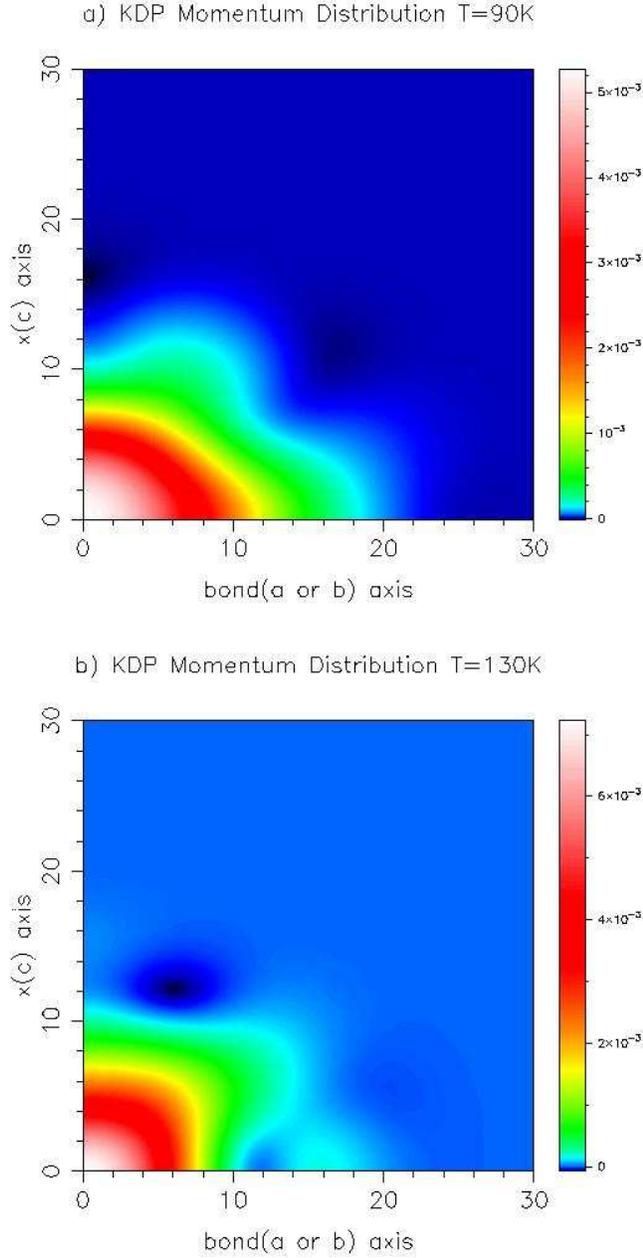,width=120mm} \caption{The
momentum distribution in the $p_x-p_z$ plane for a single hydrogen
bond below(T=90K) and just above(T=130K) the ferroelectric
transition at T=124K.  The bulge in (b) along the direction
roughly 30deg from the bond axis, that is the projection of the
displacement from the center of the bond to the phosphorus ion, is
interpreted as being due to the repulsion of the proton by the
phosphorus.} \label{fig1} \end{figure}

   The momentum distribution along the bond
axis for both temperatures is shown in Fig. 2. The dotted lines
give the uncertainty in the distribution function as a result of
the uncertainty in the measured coefficients.

\begin{figure}[h]
\hspace{.15 \hsize}\vspace{.1in}
\psfig{figure=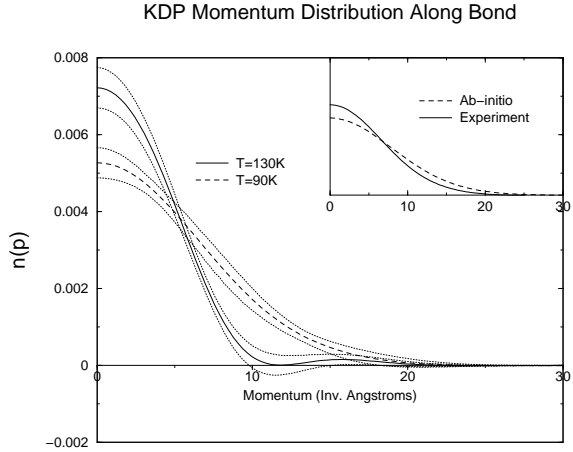,width=75mm,angle=-90} \caption{The momentum
distribution along the  bond axis for temperatures just above and
well below the structural phase transition at 124K. The dotted
curves surrounding the heavy lines are one standard deviation
error markers.   The inset is a comparison of the measured
momentum distribution at T=90K with a recent calculation by Koval,
Kohanoff and Migoni[11] in which their ab-initio one particle
potential along the bond was used to calculate the momentum
distribution from the one dimensional Schrodinger
equation,assuming the three dimensional potential was separable.}
\end{figure}   We see that overall, there is considerable
narrowing of the distribution in the high temperature phase,
indicating an increase in the length along the bond over which the
spatial wave function is coherent. The prominent feature of the
high temperature distribution, the zero and the subsequent
oscillation, is precisely what one would expect for a spatial
wave-function that was coherent over both sites, with the position
of the zero being determined by the separation of the sites.

  The momentum distribution contains many body effects due to the motion of the
surrounding ions. These effects are quite small for the heavy
ions\cite{wl}, but could be significant for the interaction of the
protons with themselves. If we  treating them in a mean field
approximation, we can calculate an effective one body
Born-Oppenheimer potential from the momentum distribution in the
high temperature phase, assuming the proton is in its ground
state\cite{rs}.   We will assume the potential is separable to
simplify this calculation,
 so that we only need the data along the $p_z$ axis.

The error bars on the measurements are such that there is
considerable uncertainty in the tail of the distribution, and one
could argue that there is no oscillation. If, however, we take the
zero of the most probable distribution to be real, and choose the
sign of the wave-function as negative for momenta greater than the
momentum at the zero , then we find the effective Born-Oppenheimer
potential and the spatial wave function shown in Fig. 3. If we do
not change the sign , we get a completely un-physical potential.
It is clear that the most likely  momentum distribution measured
supports a tunnelling model. The tunnel splitting is 94
mev($\sim$1000K), so that our assumption that we are seeing only
the ground state is consistent. Even if one doesn't take seriously
the oscillation in the  tail of the momentum distribution, and
assumes that the actual distribution goes smoothly to zero, we
would find a much broader potential above T$_c$ than below, and
the order-disorder model would be equally untenable. That
potential would, however, be inconsistent with the neutron
crystallography\cite{rn}, which clearly shows a double peak
structure for the spatial wave-function, as in Fig. 3.
\begin{figure}[h]
\hspace{.15 \hsize}\vspace{.1in}
\psfig{figure=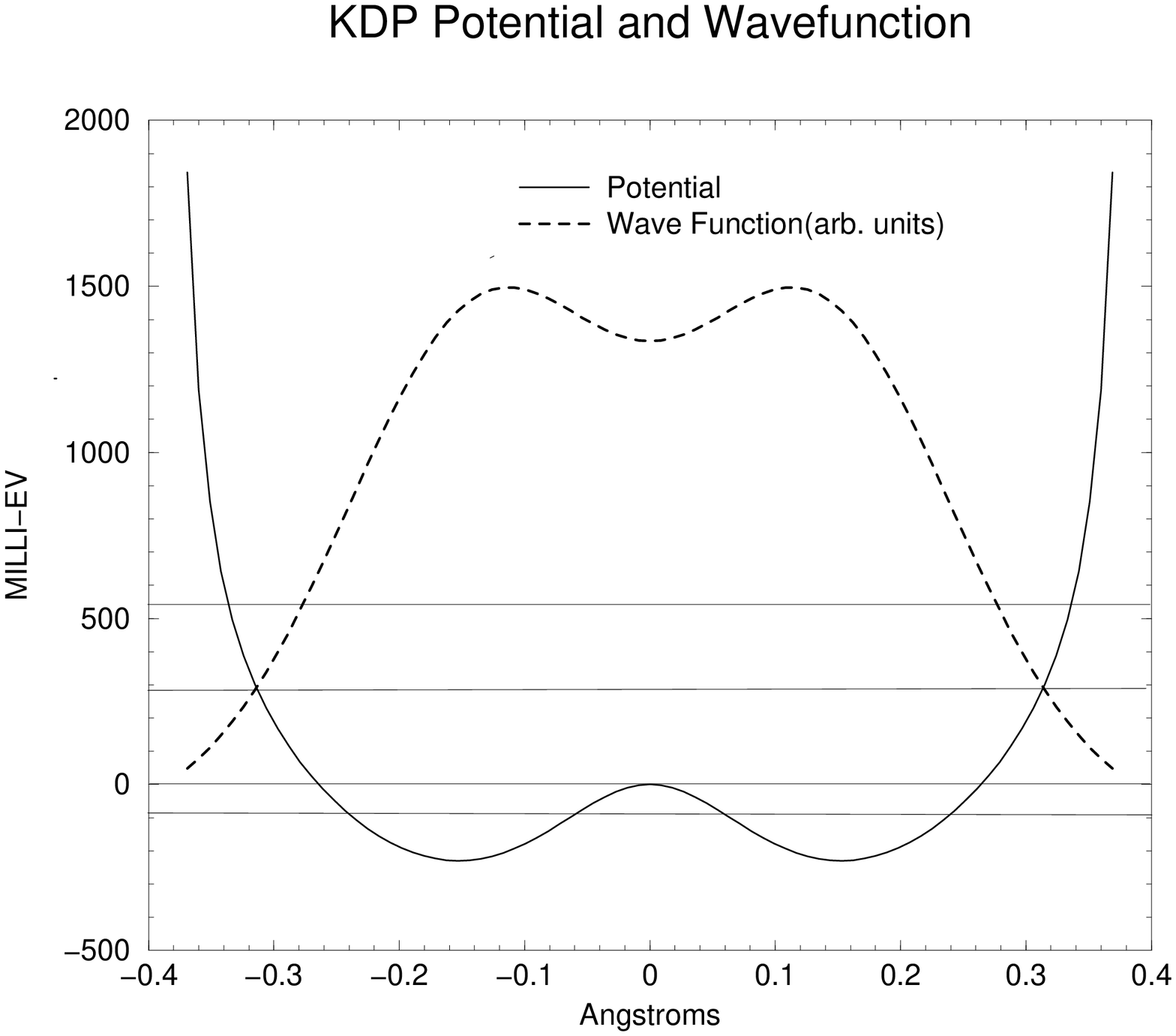,width=75mm,angle=-90} \caption{The
effective Born-Oppenheimer potential that corresponds to the
measured momentum distribution along the bond at T=130K, (Fig. 2)
, together with the spatial wave function for that potential.  The
horizontal lines give the first four energy levels for the
measured potential.}  \label{fig3}
\end{figure}
Although the potential in Fig. 3  is quite different from that
inferred from infrared absorption data by Lawrence and
Robertson\cite{lr} who assumed that it could be described by
back-to-back Morse potentials, the transition energies they based
the fit on are substantially the same as those in the figure. LR
identify the 0$\rightarrow$3 transition as having an energy of
4600 cm$^{-1}$, whereas we obtain 5088 cm$^{-1}$. For the
1$\rightarrow$2 transition, LR obtain 2260 cm$^{-1}$ while we
obtain 2039 cm$^{-1}$.
 Neutron crystallography studies have fit the shape of the spatial
wave-function with the sum of two displaced Gaussians, which would
give a rather different potential than that of Fig. 3, and
obtained a separation of the minima of the potential, of  .34
Angstroms. We find .31 Angstroms for this separation.  The shape
of the  spatial wave function in Fig. 3 is qualitatively similar
to that described in Ref. 1. We conclude that the potential of
Fig. 3 is at least
 qualitatively correct, and that the proton is indeed tunnelling
  between equivalent sites in the high temperature phase.

  Below the transition, since the potential is not symmetric, it is
not possible to invert the data to obtain it, as phase information
is lost in the momentum distribution. If one has a prediction for
the potential, one can, however,
 calculate the momentum distribution. In the inset of Fig. 2, we show the
momentum distribution obtained from a  recent\cite{kkm} ab-initio
calculation of the Born-Oppenheimer potential for KDP in which the
electronic many-body problem was treated by density functional
theory . The one particle approximation is evidently reasonably
good here. The same calculations shows that there are significant
interactions amongst the hydrogens above $T_c$, so that an
unambiguous one particle potential cannot be obtained
theoretically, as the potential for the motion of a single
particle depends sensitively on the positions assumed for the
surrounding hydrogen. The present measurements provide a means of
testing theoretical inclusions of these many body effects on the
effective one particle potential, which should reproduce the
observed momentum distribution.

If a hydrogen bond were isolated, the momentum distribution would
have to be symmetric about the bond axis. We find that
distribution in the $p_x-p_y$ plane, (not shown here) is
significantly broader at 45 degrees to either axis than it is
along the axes. This direction is approximately the projection in
the x-y plane of the vector from the center of the bond to the
phosphorus ion\cite{nm2}. If we return to Fig. 1, we see that
there is also a broadening of the distribution in the high
temperature phase at approximately 30 degrees to the z axis, which
is  the approximate direction of the projection of the vector from
the center of the bond to the phosphorus ion on the x-z plane.
This broadening becomes a narrowing at low temperatures. We
conclude that this broadening in momentum space is a result of the
repulsion of the proton by the phosphorus ion, and that the
transition displaces the particles in a way that  reduces that
repulsion. This is in broad agreement with both the calculation
cited above and another recent ab-initio calculation\cite{kio}
that showed that the motion of the phosphorus had large effects on
the hydrogen potential surface.  We conclude that the experiments
are easily sensitive enough to see the effect of the surrounding
ions on the hydrogen bond, and include many body effects that are
difficult to calculate with existing methods. This sensitivity
opens up the possibility of studying the effect of the systematic
variation of the surrounding ligands on the dynamics of hydrogen
bonds. The knowledge gained could then be used to identify the
best models and  ab-initio approximation schemes, and to infer the
behavior of the bonds in environments where the measurements can't
be done because there are two many inequivalent hydrogens, such as
in DNA or proteins. \\\\
\begin{acknowledgements} We would like to thank Devinder Sivia for assistance in
analyzing the data, Richard Nelmes, Jorge Kohanoff , Nicholas
Kioussis and Naresh Dalal for useful discussions,  and Jorge
Kohanoff for permission to use the
results of Ref. 11.
\end{acknowledgements}\\\\

\vfill\eject

\end{document}